\documentclass[%
 reprint,
 amsmath,amssymb,
 aps,
]{revtex4-1}

\usepackage{graphicx}
\usepackage{dcolumn}
\usepackage{bm}
\usepackage{upgreek}

\begin{document} 

\author{Siddharth Ghosh$^\dagger$}
\email{sg915@cam.ac.uk}
\altaffiliation{Current Affiliations: Single-Molecule Optics Group, Huygens Laboratory, Leiden Institute of Physics, Leiden University, The Netherlands.}
\altaffiliation{Centre for Misfolding Diseases, Department of Chemistry and Maxwell Centre, Cavendish Laboratory, University of Cambridge, Cambridge, UK.}

\author{Narain Karedla}
\altaffiliation{International Max Planck Research School for Physics of Biological and Complex Systems, G\"ottingen, Germany.}

\author{Ingo Gregor}
\affiliation{III. Institute of Physics -- Biophysics and Complex Systems,
  University of G\"ottingen, G\"ottingen, Germany}

\title{Single-molecule nanofluidics in a uniform dielectric confinement with resolved molecular shot noise} 

\begin{abstract}
 Until now, we could not engineer Nature's ability to dynamically handle single molecules in tunnelling-nantubes or pore-forming proteins.
 Consistent handling of individual single molecules in an extended flow will be of paramount importance to fundamental molecular studies and technological benefits, like single-molecule level separation and sorting for early biomedical diagnostics, microscopic studies of molecular interactions and electron/optical microscopy of molecules and nanomaterials. 
 However, at nanometre lenghtscales, dynamics of single molecules can only be reliably resolved if the flow is confined within a uniform dielectric environment as interacting surfaces modify electronic properties of the molecules.
This report presents an effective dynamic nanofluidic detection of optically active single molecules. 
We developed a solid-state method of nanofabricating multiplexed all-silica nanofluidic devices using shadow-angle electron-beam-deposition. 
In nanofluidic flow of single molecules, a uniform dielectric environment is essential to avoid fluctuations in electrodynamic interaction for reproducible photophysical response in dynamic single-molecule fluorescence experiments. 
The device enabled us to dynamically (electrokinetically) transport individual single molecules, like biologically relevant fluorescent probes (carbon nanodots and organic fluorophores) and small fragments DNA. 
We analysed the electrokinetic 1D molecular mass transport using two-focus fluorescence correlation spectroscopy (2fFCS) and showed confinement-induced modified molecular interactions while resolving molecular shot noise, which can misrepresent as different size or species of molecules. 
This first demonstration of high-throughput nanochannel fabrication and 2fFCS-based 1D confined detection of fast moving single molecules opens the avenue for single-molecule experiments where manipulation of dynamics is necessary, specifically for ultra-sensitive separation of biomolecules. 
\end{abstract}

\maketitle 

\section{Introduction}
Due to Brownian motion, it is impossible to keep a molecule within the detection volume for an extended period of time \cite{CohenPRL2003, CohenPNAS2006}. 
Sorting small biomolecules with single-molecule level resolution is still a challenge \cite{goodwin1993rapid, jones2017continuous, Knoll2018} even after 30 years of single-molecule detection \cite{Keller1984, Keller1987, moerner1989optical}. 
Nanofluidics \cite{Eijkel2005} can provide means to circumvent these problems for various scientific applications \cite{Noy2006, Suman2009, Noy2017}.
Nanofluidic devices comprising arrays of nanochannels with diameters less than 100\,nm have become significant for analytical investigation of DNA optical mapping \cite{Persson2010a, Kim2011, Bashir2011, Wang2015}, single virus and nanoparticle detection as well as isolation \cite{Novotny2010, Hawkins2010, Cleland2011, Merkoci2011, Sanli2015, Stolovitzky2016, Knoll2018}, ion trapping \cite{Bocquet2019}, and energy harvesting \cite{Bocquet2017}.
Nanochannels can allow systematic studies of single particles from molecules to viruses over long periods of time by suppressing the thermal motions in two directions---1D fluidic confinement. 

Even today, there is no simple and efficient way to fabricate arrays of enclosed 1D nanofluidic channels. 
Different methods have been published describing the fabrication processes of such devices.\cite{Persson2007, Bien2012, Park2010, Tegenfeldt2010, Han2005}
However, the majority of them are technically challenging, time consuming, and not easily applicable for high-throughput production.
Another problem is photophysical properties of fluophores change due to electrodynamic interactions with nonuniform semiconductor-insulator or metal-insulator interfaces.\cite{Tegenfeldt2010, vanHulst2013, vanHulst2014, Ghosh2014-1, Hao2017} 
Nonuniform dielectric nanometre confinements in dynamic experiments of single-molecule fluorescence result to misinterpretation of acquired signal or artefacts.   
In a uniform dielectric confinement, molecules can be reliably identified based on their intrinsic photophysics. 
This is not possible in environments with nonuniform electrodynamic interactions. 

Here, we present a simple nanofabrication technique to create all-silica nanochannels using electron beam lithography (EBL) and shadow-angle electron-beam-deposition (SAEBD), which are suitable for multiplexed dynamic single-molecule detection. 
The diffraction limited all-silica nanochannels were used to quantify 1D flow and diffusion of single molecules, such as small DNA molecules labelled with single organic fluorophores, carbon nanodots (CND) \cite{Ghosh2014}, and organic fluorophores by detecting their fluorescence. 
Using two-focus fluorescence correlation spectroscopy (2fFCS) \cite{Eigen1999, Joerg2007, Joerg2008, Joerg2010}, we analysed transitions of single molecules through the nanochannels and quantitatively analysed their flow velocities.
The nanofluidic detection of single molecules presented here have been preformed inside a cross-sectional diameters ranging from 30\,nm to 100\,nm. 
As far our knowledge goes, dynamic 1D manipulation and 2fFCS investigation of single molecule (of sub-3 nm sizes) inside uniform dielectric nanofluidic confinement has not been demonstrated earlier.

\section{Fabrication of nanochannel}
The process steps to create enclosed nanochannels involve fabrication of open nanochannels (trenches) using EBL and reactive ion etching (RIE). In a final step, the trenches are closed using SAEBD.
SAEBD utilises the ballistic path of the electron beam (e-beam) assisted evaporation.\cite{Becker2013}
When a collimated beam of evaporated material hits onto an open nano-trench at shallow incident angle, no deposition occurs in the shadowed region. \cite{Nakayama2007, Donald2010}
Deposition on the exposed sites causes a growth of material that can enclose a large number of parallel trenches (depending on the beam diameter) leaving the shadowed regions open as paths for fluids.
The process is unaffected by nanometre sized residues of the e-beam resists and does not require an atomically clean surface, unlike any wafer bonding-based process.\cite{Ghosh2014-1}
For photophysical reasons, the nanochannels were prepared using pure silicon dioxide (silica, SiO$_2$).
We demonstrate the SAEBD process by scanning electron microscopy (SEM) images of the intermediate steps. 
Here, we first demonstrate the process using silicon to obtain intermediate steps because silica shows low contrast in SEM due to its dielectric nature.

\subsection{Nano-trenches}
Figure~1\textbf{a} shows a schematic flow-chart of creating nanochannels on Si-[100] wafers. The width of the nano-trenches (i.e. the width of the final nanochannels) can be optimised by the e-beam exposure of the EBL to the positive resist. 
Different widths of nano-trenches were created ranging from 30\,nm to 100\,nm. 
In Figure~1\textbf{b}, 65\,nm and 100\,nm wide nano-trenches are shown. 
The lithographed e-beam resist acted as a mask for RIE to etch the final nano-trenches on silicon.
We checked the depth of the trenches using AFM. For 40\,nm wide trenches we found depths in the range of 35\,nm to 40\,nm (supplementary Figure~1).

\subsection{Nanochannels -- enclosing nano-trenches}
In the next step, SAEBD was used to enclose the nano-trenches to create closed nanochannels.
Figure~1\textbf{c} schematically explains the concept of the SAEBD process: a high-energy e-beam is bent by a magnet onto the reservoir of material that sublimates and finally deposits onto the substrate.
Figure~1\textbf{d} schematically explains the role of the deposition angle ($\theta$) between the surface of the substrate and the vapour. The angular deposition creates a shadowed region that is unexposed to the depositing material.
The deposited and shadowed regions are colour coded in red and yellow, respectively.
To demonstrate this, an array of 5\,mm long nano-trenches were cross-sectioned using a wafer sawing instrument to observe the intermediate steps while performing SAEBD.
Figure~1\textbf{e} schematically represents a high angle deposition which was experimentally performed at $80^\circ$ as shown in Figure~1\textbf{f}.
As shown in the time evolution schematic, SAEBD closes the nano-trenches leaving apart a void. By decreasing $\theta$ the unexposed area increases. At an acute angle close to $0^\circ$, the growth is nearly parallel to the surface of the substrate as shown in Figure~1\textbf{e}-\textbf{h}.
Figure~1\textbf{g} schematically represents a low angle deposition and experimental demonstration of that is shown in Figure~1\textbf{f} at $45^\circ$. Due to instrumental constraints, it was not possible to achieve a deposition at an acute angle close to $0^\circ$.
Nevertheless, satisfactory results were obtained using $\theta = 45^\circ$ as shown in Figure~1\textbf{i}-\textbf{k}.
Here, 60\,nm titanium was deposited on the open nano-trenches at an angle of $45^\circ$ with a deposition rate of 1\AA/s at a pressure of $2\times 10^{-6}$\,mbar.
We estimate that at an angle $30^\circ$ a high-quality flat edge will be formed.

We used a focused ion beam (FIB) to investigate the cross-sections of the nanochannels. To avoid ion beam induced damage in FIB, the top part of the cutting region was protected with thin metallic layers.
We deposited two thin-films on the top surface of the enclosed nanochannels.
Figure~1\textbf{i} - \textbf{k} show SEM images of the milled regions from low to high magnification. In Figure~1\textbf{j}, the first layer (Si-1) is the silicon substrate on which nano-trenches were fabricated.
The second layer is titanium (Ti-2)  that was deposited under $45^\circ$ SAEBD.
Layers Pt-3 and Pt-4 are platinum layers of 100\,nm and 450\,nm, respectively, which acted as protective layers to avoid FIB induced damage.
In Figure~1\textbf{k}, we observe the magnified cross-section of an enclosed nanochannel.
As expected, SAEBD growth of titanium produces a well defined flat layer closing the nanochannel. The vertical thickness of the titanium film is 47.8\,nm, This corresponds well with the experimental settings, that were set to deposition of 60\,nm Ti under an angle of $\theta=45^\circ$ that should result in a layer of ~51\,nm vertical thickness.

After this proof of concept we prepared fused silica based nanofluidic devices.
Here, 5\,nm of gold film was sputter-coated prior to spin-coating the e-beam resist on the silica wafer. This is necessary to reduce the charging effect of silica under EBL. 
After RIE, the obtained nano-trenches in the silica wafer were enclosed with SAEBD using silicon dioxide under $45^\circ$.

\section{Results and Discussion}
\subsection{Nanofluidic device and 2fFCS detection scheme}
The design of the silica based nanofluidic device to perform single molecule experiments is schematically illustrated in Figure~2\textbf{a}.
Two reservoirs with a diameter of a few millimetres were sand blasted on the silica wafers using $70\,\upmu$m silica particles prior to the nano-fabrication process. They are separated by a distance of about one millimetre and serve as convenient macroscopic in- and outlet for the fluids. Each of these reservoir is connected to an array of parallel ~30 micrometre wide channels. These channels lead to either side of the final array of $<100$\,nm wide channels. These nanochannels span over a length of $l = 200\,\upmu$m and connect the microchannels. The height of the nano- and the microchannels are the same and are etched in the same step using RIE.

In Figure~2{a}, the white and blue regions indicate the reservoir and the microchannels, respectively (supplementary Figs~2, 3, and 4).
The red stripes correspond to the array of nanochannels, that are connected to the microchannels. A SEM image of these silica nanochannels is shown in the right inset of Figure~2\textbf{a}.
For the experiment, we filled the reservoirs with a dilute solution of fluorescent probes in buffer. After filling one reservoir capillary force transported the fluid through the microchannels and into the nanochannels reservoir. A relaxation time of 30\,s was given to avoid development of trapped air bubbles between two inlets. Then, the second reservoir was filled. Two 100\,${\upmu}$m thick platinum electrodes were immersed into the reservoirs (supplementary Figure~5), and an electric field was applied along the nanochannel. This created an electro-osmotic flow \cite{Zare1988, Hu1998, Zhang2000, Suman2007} that uni-directionally transported the fluid through the nanochannels as shown in Figure~2\textbf{b}.

To restrict unwanted surface adsorption, we used fluorescent molecules carrying the same charge as the nanochannels' wall.
Pure silica is negatively charged above its isoelectric point (pH(I) = 2).\cite{Kosmulski2001, Raghavan2009}
The buffer's pH of 8.5 leads to a considerable amount of negative charges on the walls of silica nanochannels.
This condition efficiently reduces adsorption of negatively charged molecules in the channels.

Figure~2\textbf{c} shows single AlexaFluor~647 molecules (Thermo Fisher, Massachusetts, USA) lined up horizontally in all the parallel nanochannels. To immobilise the molecules, the solvent was dried at room temperature leaving behind the molecules in the channels. The  photon counts profile shows an average SNR of 90. The image was captured using a wide-field optical microscope by exciting the molecules with a 640\,nm cw-laser (Coherent Laser Systems GmbH, G\"ottingen, Germany).
Beside the evidence of FIB and SEM, this also proves that the nanochannels are properly enclosed and no cross-links between nanochannels are observed, unlike our previous observation.~\cite{Ghosh2014-1}

We used 2fFCS to study the dynamics of the molecules inside the nanochannels. The diffraction limited focus of a laser beam in the visible range has a typical diameter of $D_f \approx$ 300\,nm to 500\,nm. This is much larger than the width of the nanochannels, which is in the range of $d_{nc} \approx$ 30\,nm to 100\,nm. Under this condition only a movement along the channel can be detected. Therefore, we can consider the flow inside a nanochannel as \textsl{quasi} 1D. (Figure~2\textbf{a}).
In 2fFCS two laser foci are used that are pulsed alternately with a rate of 20\,MHz. By performing temporal correlation of the signals from both foci one can accurately determine the times that a fluorescent entity takes to move from one focus to the other. In our experiment, the displacement of the foci was $\approx$ 400\,nm and they were carefully aligned with the direction of the nanochannel (Figure~2\textbf{b}).~\cite{Joerg2007}
A \textit{x-y-z} piezo-scanner was used to move the device to the position of the foci and allows to acquire point measurements as well as confocal scan images.

\subsection{APD based detection}
As the first probe of interest, we chose $<$2\,nm sized carbon nanodots\cite{Ghosh2014} which were later renamed as graphene quantum dots (GQD)\cite{Ghosh2016} and are negatively charged (see supplementary Figure~6). Figure~3\textbf{a} shows a schematic top-view of the device, where the nanochannels appear as dark lines.
A confocal scanned image of the \textit{x-y} plane was recorded, while GQDs were flowing inside the nanochannels (Figure~3\textbf{b}). The applied electric field to induce electro-osmotic flow during the measurement was 15\,V/mm.
The pixel size of the image is 320\,nm with dwell time of 5\,ms per pixel. 
Figure~3\textbf{c} shows a scheme of a \textit{y-z} section through the nanochannels. 
In Figure~3\textbf{d} to \textbf{g} we present a series of recorded \textit{y-z} scanned confocal images. Here, the volume percentage of GQDs relative to the stock solution increases from $1\%, 2\%, 5\%$, to $50\%$, respectively. In these scans the pixel size was 100\,nm with a dwell time of 2\,ms per pixel.
In these scans, we observe several periodic point spread functions (PSF) as the fluorescence is emitted from sub-diffraction sized volumes. At low concentrations of GQDs the fluorescent signals are strongly fluctuating due to the motion of the fluophores out of the excitation focus. As the concentration of GQD increased, the fluctuations of the photon signal decreased and a clear images of the PFSs are obtained. 
Since the GQDs were close to size of organic fluorophores, it could not have been possible to measure the fast flow in a wide-field emCCD-based setup (supplementary Figure~7 and section emCCD-based detection). 

\subsection{Electro-osmotic flow of single molecules}
\subsubsection{2fFCS}
We investigated the flow velocities of single-molecules in the nanochannnels using 2fFCS. In particular, we used the organic dye Atto~488 (Atto-Tec GmbH, Siegen, Germany) and 48 base-pair dsDNA labelled with AlexaFluor~647 (IBA GmbH, G\"ottingen, Germany). The molecules were diluted to a concentration of about 1\,nM in aqueous buffer solution (see 'Methods' section). We measured the flow velocities and diffusion of single molecules inside nanochannels having a width of 30\,nm. The applied electric fields range from 27\,V/mm to 300\,V/mm (see supplementary Figure~9). 
The temporal cross-correlation functions of the photon traces between the two excitation foci together with the well known distance of the foci allow to accurately determine the flow velocity of the fluorescent probes.\cite{Joerg2008, Joerg2010}.
The correlated data points from 2fFCS were fitted with the Fokker-Planck equation \cite{Risken1984} considering the 1D electro-osmotic flow.
A challenge for these measurements is the tight spatial confinement of the molecules not only in the \textit{z} plane, but also in the \textit{y} axis. To find the optimal position for the measurements, we recorded a \textit{y-z} confocal scan around the selected nanochannel. The point that showed the highest number of photons in the scan was then chosen as point of measurements. 

An exemplary 2fFCS measurement performed at 220\,V/mm is shown in Figure~4\textbf{a}.
Here, blue and red curves are fits to the two cross-correlation data points. The other two are fits to the autocorrelation data, respectively.
The velocity and diffusion coefficient from the curve fitting are $D = 1.51 \times 10^{-7} \mathrm{cm^2/s}$ and $v=-207 \upmu$m/s.
The negative value of the flow velocity infers the direction of the flow from focus 2 to focus 1, which can be altered by changing the applied polarity of the electric field.
The linear relationship of electro-osmotic flows at different applied electric fields is plotted in Figure~4\textbf{b}. The linear fit has an $r^2$ value of $0.992$. A fluctuation of diffusion coefficient is observed in the range of $1.1 \times 10^{-7}$ to $9.6 \times 10^{-7}$ is observed (supplementary Figure~10). However, these values are not very accurate under these conditions, since the driven flow clearly dominates the dynamics of the molecules. 

\subsubsection{Photon burst size distribution}
To confirm that the molecules do not from aggregates under the conditions in the confined nanochannels we analysed the photon burst sizes of the molecular transits.\cite{Joerg1998}. Figure~4\textbf{c} shows the burst size distribution (BSD) of the time trace of a flow measurement. The time trace is shown in the inset.
The first peak with the lowest photon number is due to the single molecule transits. The second and higher peaks are due to two and more molecules coinciding in the focus. If the molecules move independently as single entities, the BSD is given by a Poissonian distribution. As the fitted curve shows, for the given data this is clearly the case. Figure 4\textbf{d} shows the first 600\,ms of the inset 5\,ms binned time-trace of Figure~4\textbf{c}. This also shows clearly bursts of single molecule transits.

\subsubsection{Simulation of confined diffusion}
For theoretical interpretation, we simulated a similar confined nanofluidic molecular interactions in LAMMPS platform \cite{Plimpton1995} (Figure~4\textbf{e}, also see the supplementary movie S3) where the walls of the nanochannel and particles have same charge. 
The kymographic analysis (Figure~4\textbf{f-h}) of the three different regions inside the nanochannel shows characteristic variations near the walls and at the centre of the wall. 
Along the centre line of the nanochannel discrete single events are observable along with successive multiple-molecule-events (highlighted in green in Figure~4f). 
Molecular interactions near the wall (Figure~4\textbf{g}) are dominated by a large number of crawling events where molecules were moving along the wall for extended periods of time -- ranging from 10 to 130 frames. These events are highlighted in red. 
In Figure~4h, at the wall -- fluid-solid interface, both the red and green events are not dominant. 
Within the diffraction limited confocal detection volumes of 2fFCS, besides the molecular shot noise, the crawling events are another potential cause behind the second less frequent peak in the BSD analysis and variation in diffusion coefficients.
Surface chemistry -- a function of surface roughness and ionic contaminants play an important role in electrokinetic transports at nanometre length scale as observed by Golestanian and co-workers \cite{Golestanian2014}.

\section{Conclusion}
At the length scale where molecular interactions can be strongly influenced by surface charge, a uniform dielectric confinement removes unwanted artefacts generated by non-uniform electrodynamic interactions. 
The observed molecular behaviour in the fabricated all-silica nanofluidic channels, demonstrates absence of confinement induced artefacts.
In summary, this Letter presents an efficient method to fabricate all-silica nanochannels using SAEBD and their usage to reduce thermal motion in two spatial dimensions in single molecule experiments. 
The nanofabrication process demonstrated here is suitable for high-throughput and large scale production of nanofluidic chips. SAEBD is not material restricted, unlike the oxidation and bonding based techniques. 
Large range velocity variations of 1D flow of single organic fluorophores, sub-2\,nm GQDs, and 11\,nm DNA molecules were achieved using electro-osmosis inside nanochannels. As far our knowledge goes, this is the first quantitative study of 1D nanofluidic flows in nanochannels at single-molecule level utilising 2fFCS. 
A broad range of velocities of up to 300\,$\upmu$m/s were achieved by varying the applied electric field. 
BSD analysis confirms that the observed transits were mainly due to single emitters. All the experiments were performed inside nanochannels with a cross-section of 30\,nm $\times$ 35\,nm. The experimental observations are justified with BSD analysis and MD simulation. 
This simple approach of fabricating nanochannels paves the way towards detecting early onset of any disease at single molecule level. 
In future, trapping nanoscale objects of less than 2\,nm in size for large residence time should be also feasible using these nanochannels. 
Biomolecular interactions such as dynamics of DNA, protein aggregation, and structural biology of molecules under physiological condition can be also studied at single molecule level using the SAEBD based nanofluidic devices. 
The fundamental understanding diffusion inside nanofludic channel requires further investigation and opens another path of fundamental research.

\section*{Acknowledgement}
SG thanks the International Max Planck Research School for Physics of Biological and Complex Systems and the Ministry of Science and Culture (Lower Saxony) for awarding an Excellence Stipend/MWK PhD scholarship. 
The work presented here was carried out at the research group of Prof J\"org Enderlein. 
The authors thank the internal funding from the University of G\"ottingen associated to the Third Institute of Physics for this project. 
Lastly, the authors are extremely grateful to Prof Enderlein for providing plenty of scientific advice. 
Dr Mitja Platen and Ditter Hille has supported us in AFM and building the SAEBD stages, respecitively. 
Authors also thank Xuxing Lu (Single Molecule Optics Group, Leiden) for suggestions on the simulations. 

\section*{Authors Contribution}
SG has written the manuscript, conceptualised the idea of nanofabrication, prepared the nanofluidic device, performed all the measurements and numerical fitting presented in this paper. SG and NK have performed initial measurements together. NK has modified the fitting routine provided by Prof J\"org Enderlein and IG. IG has supervised the project and provided scientific advice. All the authors have read and made adequate corrections in the manuscripts.

\section*{Competing interests}
The authors declare no competing interests.

\section*{Additional information}
The supplementary information contains the methods of nanofabrication, figures and design of the nanofludic device, integrating the nanofluidic device to electrodes, carbon nanodots/GQD cation-pi interaction behaviour, EMCCD-based measurements, description of 2fFCS setup, 2fFCS fitting function, control 1D diffusion of single Atto 488 fluorophores, electro-osmotic flow and diffusion of single DNA molecules at different electric fields, and LAMMPS simulation. \\

\bibliography{Ref}

\clearpage

\begin{figure*}[b]
\includegraphics[scale=0.55]{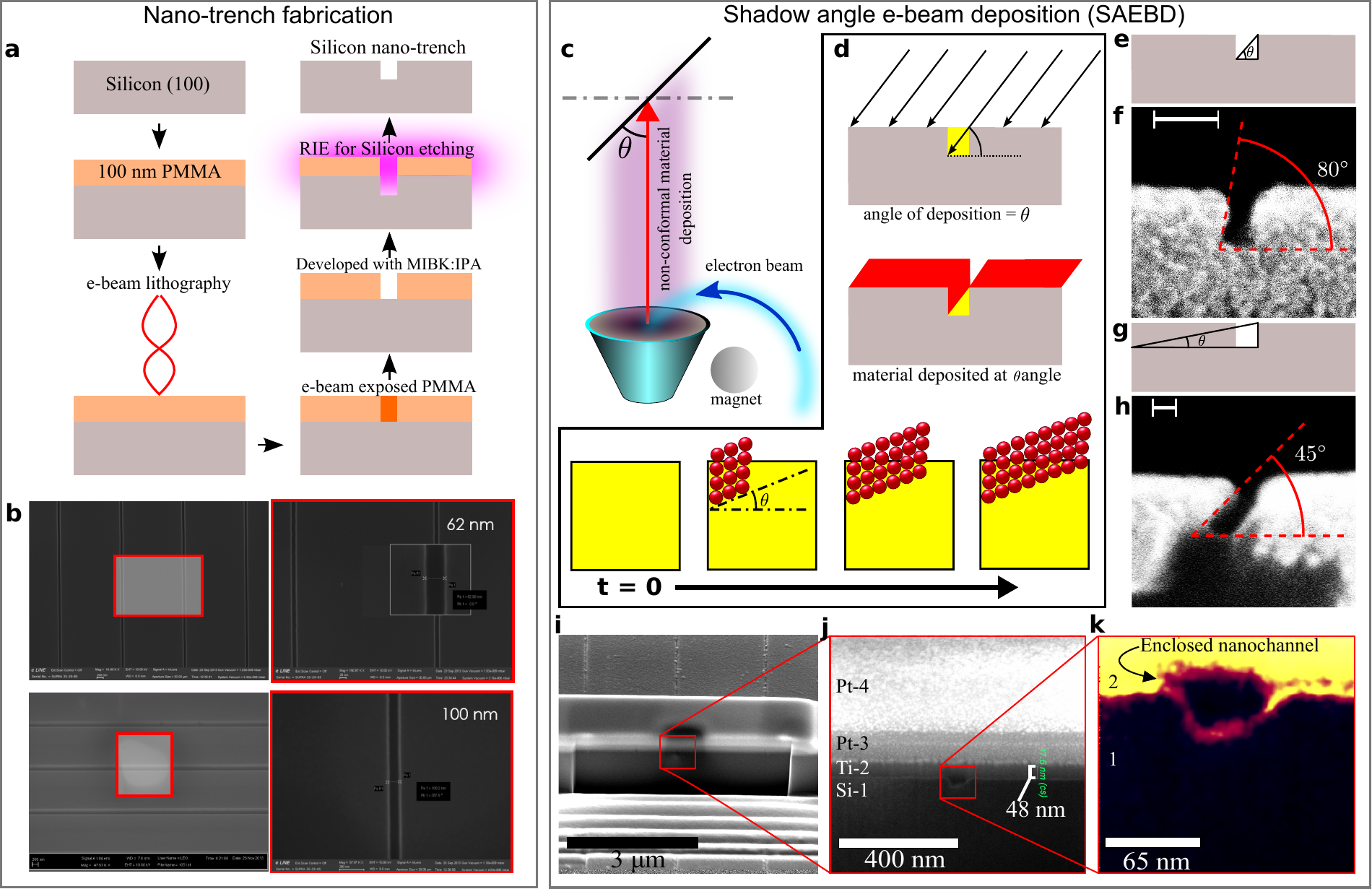}
\caption{\textbf{Nanochannel fabrication using SAEBD.}
\textbf{a.}~Fabrication of nano-trenches on silicon with EBL and RIE.
\textbf{b.}~SEM of silicon nano-trenches with 62\,nm and 100\,nm width.
\textbf{c.}~SAEBD at angle $\theta$.
\textbf{d.}~Shadows of electron beam. Arrows indicate angular e-beam evaporation.
\textbf{e.}~Schematic of high angle deposition and \textbf{f.} SEM of deposition at $80^\circ$.
\textbf{g.}~Schematic of low angle deposition and
\textbf{h.}~SEM of deposition at $45^\circ$.
\textbf{i.}~FIB cross-section of the enclosed nanochannels.
Two layers of platinum were used to protect the nanochannel form high energy ions.
\textbf{j.}~Magnified view of the enclosed nanochannel, where Si-1 is the silicon substrate on which nano-trenches were fabricated, Ti-2 is the 50\,nm thick titanium layer which was used in SAEBD, Pt-3 and Pt-4 are platinum deposited inside FIB.
\textbf{k.}~Further magnified view of the nanochannel where region 1 is  silicon (Si-1) and region 2 is titanium (Ti-2).}
\end{figure*}

\begin{figure}[!ht]
\begin{center}
\includegraphics[scale= 0.28]{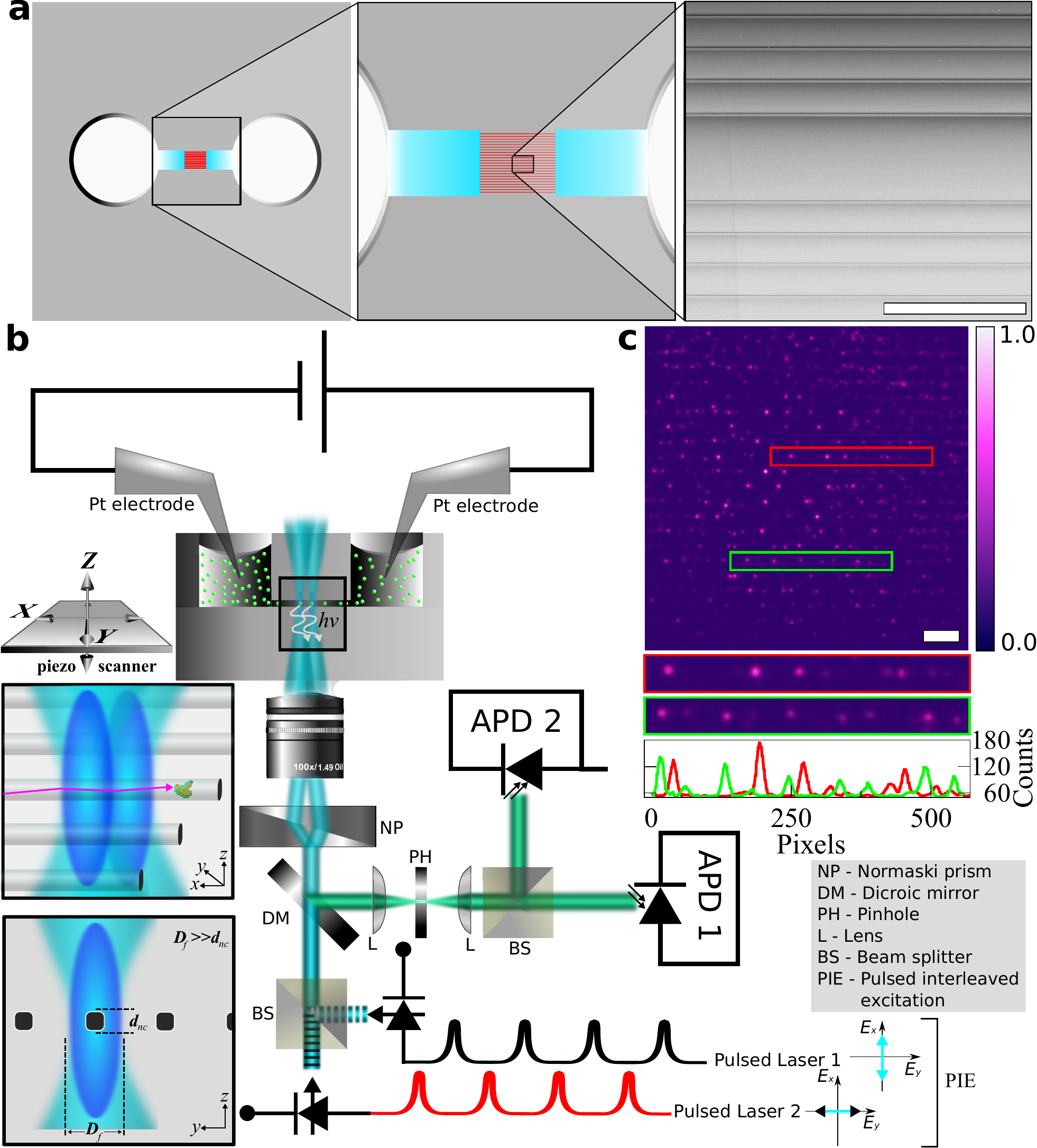}
\end{center}
\caption{\textbf{Nanofluidic device in a 2fFCS setup.}~\textbf{a.}~Schematic top view of nanofluidic devices  with an SEM image of real nanochannels (scale bar is 30 ${\upmu}$m). \textbf{b.}~Side view schematic of the complete experimental set-up where electric field is applied through two reservoirs along the nanochannels using platinum electrodes.
Two foci (of 2fFCS) were aligned with nanochannel using a 100$\times$ 1.49 NA oil immersion objective lens.
Two different linearly polarised pulsed interleaved lasers were used for in two-focus excitation.
The emission from flowing single molecules was detected by two APDs.
\textbf{c.}~A wide-field of image frame showing the presence of single molecule (scale bar is 8\,${\upmu}$m).}
\end{figure}

\begin{figure}[!ht]
\begin{center}
\includegraphics[scale=0.5]{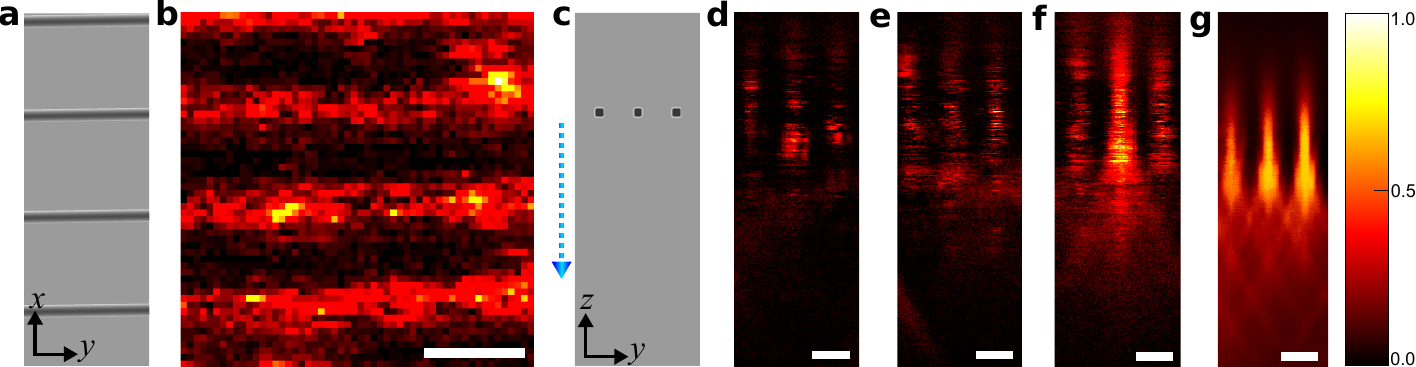}
\caption {\textbf{1D Flow of GQDs.}~\textbf{a.} Schematic top-view of nanochannels along which an (\textit{y-x}) scan was performed.
\textbf{b.}~Confocal scan image of nanochannels filled with GQDs.
\textbf{c.}~Schematic cross-sectional side view of nanochannels along which (\textit{y-z}) scans were performed. The dashed arrow represent the optical excitation path -- direction from immersion oil of objective lens to nanofluidic device.
\textbf{d}, \textbf{e}, and \textbf{f} \textit{y-z} scan images of nanochannels with an increasing order of GQDs' concentration flowing through nanochannels.
\textbf{g}.~High concentration of GQD. All the horizontal scale bars denotes 2$\upmu$m. The vertical scale bar denotes photon counts as 0.0 (lowest) to 1.0 (highest). }
\end{center}
\end{figure}

\begin{figure*}[!ht]
\begin{center}
\includegraphics[scale=1]{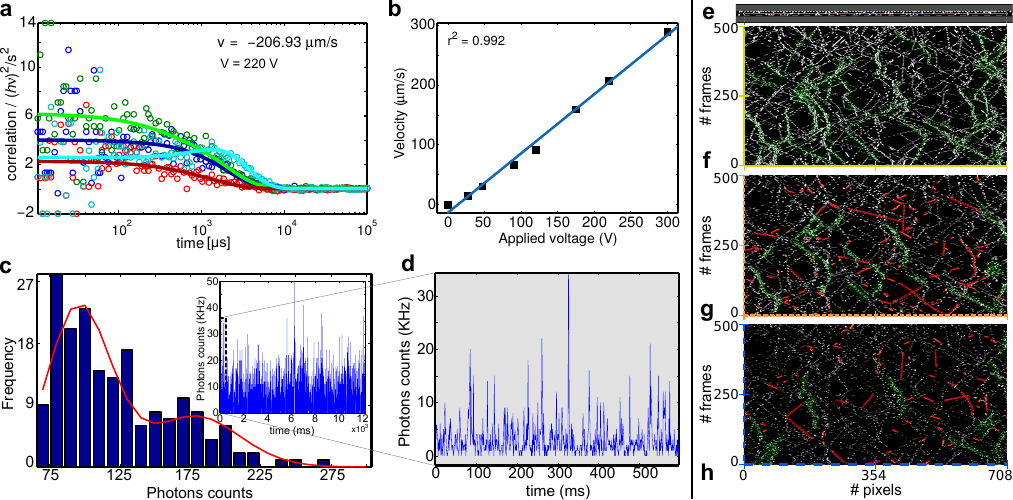}
\caption{\textbf{1D Flow of 48bp DNA.}~\textbf{a.}~2fFCS measured correlation plots of photon counts from two APDs. Here, applied electric field was of 220 V/mm. The data was fitted with 1D Fokker-Planck equation. From the fitting, we found diffussion coefficient of $1.51\times 10^{-7}$ cm$^2$/s.
\textbf{b.}~Plot of velocity versus applied voltage along the nanochannel. The data was fitted with linear fit.
\textbf{c.}~Burst size distribution of single molecule transits fitted with two Poissonian distributions. Total time trace binned with 5 ms is shown in the inset -- the dashed bordered grey region is shown in \textbf{d}.
\textbf{d.}~A part of the time trace plots of single photon bursts due single DNA molecules transits through focus. 
\textbf{MD simulation of 1D diffusion.} 
\textbf{e.}~A frame from the simulation of diffusing particles inside a confined channel. 
Kymograph \textbf{f.}~along the centre of the channel (yellow line) shows presence of molecular shot noise (green highlights), 
\textbf{g.}~along the orange dashed line i.e. close to the wall with large degree of wall interaction induced drag or crawling (red highlights), and 
\textbf{h.}~along the blue dashed line at the wall-particle interface with less degree of wall interaction in contrast to \textbf{g}.}
\end{center}
\end{figure*}

\end{document}